# Motor-driven advection competes with crowding to drive spatiotemporally heterogeneous transport in cytoskeleton composites

Janet Y Sheung[1,2*], Jonathan Garamella[3], Stella K Kahl[1], Brian Y Lee [2], Ryan J McGorty[3], Rae M Robertson-Anderson[3]

[1]W. M. Keck Science Department, Scripps College, Claremont, CA, USA
[2]W. M. Keck Science Department, Claremont McKenna College, Claremont, CA, USA
[3]Physics and Biophysics Department, University of San Diego, San Diego, CA, USA

**\* Correspondence:** jsheung@scrippscollege.edu

## ABSTRACT

The cytoskeleton–a composite network of biopolymers, molecular motors, and associated binding proteins–is a paradigmatic example of active matter. Particle transport through the cytoskeleton can range from anomalous and heterogeneous subdiffusion to superdiffusion and advection. Yet, recapitulating and understanding these properties–ubiquitous to the cytoskeleton and other out-of-equilibrium soft matter systems–remains challenging. Here, we combine light sheet microscopy with differential dynamic microscopy and singe-particle tracking to elucidate anomalous and advective transport in actomyosin-microtubule composites. We show that particles exhibit multi-mode transport that transitions from pronounced subdiffusion to superdiffusion at tunable crossover timescales. Surprisingly, while higher actomyosin content enhances superdiffusivity, it also markedly increases the degree of subdiffusion at short timescales and generally slows transport. Corresponding displacement distributions display unique combinations of non-Gaussianity, asymmetry, and non-zero modes, indicative of directed advection coupled with caged diffusion and hopping. At larger spatiotemporal scales, particles undergo superdiffusion which generally increases with actomyosin content, in contrast to normal, yet faster, diffusion without actomyosin. Our specific results shed important new light on the interplay between non-equilibrium processes, crowding and heterogeneity in active cytoskeletal systems. More generally, our approach is broadly applicable to active matter systems to elucidate transport and dynamics across scales.

## 1 INTRODUCTION

The cytoplasm is a crowded, heterogeneous, out-of-equilibrium material through which macromolecules and vesicles traverse to perform critical cellular processes such as mitosis, endocytosis, migration, and regeneration[1–4]. Macromolecules and particles diffusing through the cytoplasm and other similar materials have been shown to exhibit widely varying and poorly understood anomalous transport properties that deviate significantly from normal Brownian diffusion. In particular, the mean-squared displacement, $MSD$, often does not scale linearly with lag time $\Delta t$, but is instead better described by $MSD \sim \Delta t^\alpha$ where $\alpha < 1$ or $\alpha > 1$ for subdiffusion or superdiffusion, respectively. The distributions of displacements (i.e., van Hove distributions) also often deviate from Gaussianity and can display exponential tails at large displacements[5–8]. The cytoskeleton–an active composite of filamentous proteins including actin, microtubules, and intermediate filaments, along with their associated motor proteins–plays a key role in these observed anomalous transport properties[9–11]. Such anomalous transport phenomena are not just observed in cytoskeleton, but are ubiquitous in numerous other active and crowded soft matter systems, making their characterization and understanding of broad interest.

In steady-state, the thermal transport of particles through in vitro cytoskeletal systems exhibit varying degrees of subdiffusion and non-Gaussianity depending on the types and concentrations of filaments and crosslinking proteins[5,6,12,13]. For example, single-particle tracking (SPT) of particles in composites of entangled actin filaments and microtubules, has revealed increasing degrees of subdiffusion ($\alpha$ decreasing from ~0.95 to ~0.58) as the molar ratio of semiflexible actin filaments to rigid microtubules increased[6]. The corresponding SPT van Hove distributions were reported to be non-Gaussian, displaying larger than expected probabilities for very small and large displacements, indicative of particles being caged in the filament mesh and hopping between cages.

Differential dynamic microscopy (DDM), which uses Fourier-space analysis to measure the timescales over which particle density fluctuations decay, has also been used to measure transport and quantify anomalous characteristics over larger spatiotemporal scales compared to SPT[5,14,15]. DDM analysis of a time-series of images provides a characteristic decay time $\tau$ as a function of the wave vector $q$ which typically follows power-law scaling $\tau(q) \sim q^{-\beta}$ [15,16], with $\beta$ relating to the anomalous scaling exponent $\alpha$ via $\beta = 2/\alpha$. Specifically, $\beta = 2, > 2, < 2$ and 1 correspond to diffusive, subdiffusive, superdiffusive, and ballistic motion (Figure 1F). DDM analysis of actin-microtubule composites corroborated the SPT results described above, with subdiffusive $\beta$ values tracking with $\alpha$ values[5,6].

Similar SPT and DDM experiments demonstrated that crosslinking of actin and/or microtubules introduced bi-phasic transport with the subdiffusive scaling exponents dropping from $\alpha \approx 0.5 - 0.7$ to $\alpha \approx 0.25 - 0.4$ (depending on crosslinker type) after $\Delta t \approx 3$ s, due to strong caging and reduced thermal fluctuations of filaments. At the same time, van Hove distributions were well fit to a sum of a Gaussian and exponential, and the non-Gaussianity parameter increased, indicating enhanced heterogeneity[5,6,8,12,17].

Numerous studies have also investigated transport in non-equilibrium cytoskeleton networks, in which activity is introduced via motor proteins, such as actin-associated myosin II and microtubule-associated kinesin[18–20,10,21,2]. These studies have shown evidence of vesicle movement strongly tracking with actin movement, microtubule-dependent flow, and the simultaneous presence of subdiffusive and ballistic transport dynamics. While the majority of these active matter studies have been on systems of either actin or microtubules, recent studies have used DDM and optical tweezers microrheology to characterize the dynamics of actin-microtubule composites pushed out-of-equilibrium by myosin II minifilaments straining actin filaments[14,22,23]. These studies showed that active actin-microtubule composites exhibited ballistic-like ($\alpha \approx 2$) contractile motion, rather than randomly-oriented diffusion or subdiffusion, with speeds that increased with increasing fraction of actin in the composites, due to increased composite flexibility[14,23]. Myosin-driven contractile motion and restructuring was also reported to increase viscoelastic moduli and relaxation timescales and induce clustering and increased heterogeneity of the initially uniform meshwork [22].

However, particle transport through active actin-microtubule composites–likely dictated by the complex combination of active processes, crowding, and interactions between mechanically distinct filaments – has remained largely unexplored. The majority of studies that have examined the combined effect of activity and crowding have been in vivo[1,24–29], where a large number of conflating variables that are difficult to tune make teasing out the effect of each contribution and mechanism highly non-trivial.

Here, we take advantage of the tunability of in vitro reconstituted cytoskeleton composites to systematically investigate the coupled effects of non-equilibrium activity, crowding, and heterogeneity on particle transport. We combine fluorescence light sheet microscopy (fLSM) with single-particle tracking (SPT) and differential dynamic microscopy (DDM) to examine the anomalous transport of micron-sized particles within active composites of myosin II minifilaments, actin filaments, and microtubules with varying molar fractions of actin and tubulin (Fig 1A). We leverage the optical sectioning and low excitation power of fLSM (Fig 1B) to capture particle trajectories with a temporal resolution of 0.1s for durations up to 400 s (Fig 1C). Using both SPT and DDM provides transport characterization over a broad spatiotemporal range that spans ~$10^{-1}$ -$10^2$ s and ~$10^{-1}$ - 10 μm. From measured SPT trajectories, we compute ensemble-averaged $MSD$s and associated anomalous scaling exponents $\alpha$ (Fig 1D), as well as corresponding distributions of particle displacements, i.e., van Hove distributions (Fig 1E), for varying lag times $\Delta t$. To expand the spatiotemporal range of our measurements and provide an independent measure of transport characteristics, we use DDM to analyze particle density fluctuations in Fourier space, and evaluate the power-law dependence of decorrelation times $\tau$ on wave vector $q$, i.e., $\tau(q) \sim q^{-\beta}$ (Figure 1F).

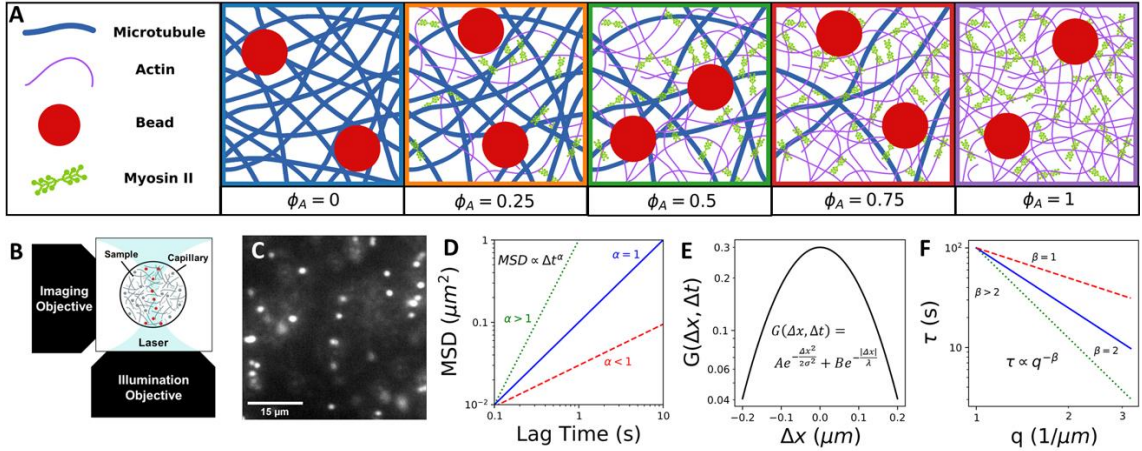

**Figure 1: Combining light sheet microscopy with real-space single-particle tracking (SPT) and reciprocal-space differential dynamic microscopy (DDM) to characterize particle transport in active cytoskeletal composites.** (**A**) We create composites of co-entangled microtubules (blue) and actin filaments (purple) driven out-of-equilibrium by myosin II minifilaments (green). We track the motion of embedded 1 μm beads (red) in composites with varying molar fractions of actomyosin, which we denote by the fraction of actin comprising the combined molar concentration of actin and tubulin (5.8 μM): $\phi_A = 0, 0.25, 0.5, 0.75, 1$. In all cases, the molar ratio of myosin to actin is fixed at 0.08. (**B**) Schematic of the light-sheet microscope we use for data collection, which provides the necessary optical sectioning to capture dynamics in dense three-dimensional samples. (**C**) Example frame from time-series of 1 μm beads embedded in a cytoskeleton composite, used to characterize particle transport in active crowded systems. (**D**) Cartoon of expected mean-squared displacements ($MSD$) of embedded particles versus lag time $\Delta t$, which we compute via single-particle tracking (SPT) and fit to a power law $MSD \propto \Delta t^\alpha$ to determine the extent to which particles exhibit normal Brownian diffusion ($\alpha = 1$, blue), subdiffusion ($\alpha < 1$, red), or superdiffusion ($\alpha > 1$, green). (**E**) Cartoon van Hove distribution $G$ of particle displacements $\Delta x$ for a given lag time $\Delta t$ computed from SPT trajectories. The distribution shown is described by a sum of a Gaussian and exponential function $G(\Delta x, \Delta t) = Ae^{-\frac{\Delta x^2}{2\sigma^2}} + Be^{-\frac{|\Delta x|}{\lambda}}$, as is often seen in crowded and confined systems and those that display heterogeneous transport. (**F**) Cartoon of expected characteristic decorrelation times $\tau(q)$ as a function of wave number $q$, which we compute by fitting the image structure function computed from DDM analysis. We determine the scaling exponent $\beta$ from the power-law $\tau(q) \sim q^{-\beta}$ to determine if transport is diffusive ($\beta = 2$, blue), subdiffusive ($\beta > 2$, green), or ballistic ($\beta = 1$, red).

## 2 MATERIALS AND METHODS

**Protein Preparation:** We reconstitute rabbit skeletal actin (Cytoskeleton, Inc. AKL99) to 2 mg/ml in 5 mM Tris–HCl (pH 8.0), 0.2 mM CaCl$_2$, 0.2 mM ATP, 5% (w/v) sucrose, and 1% (w/v) dextran; porcine brain tubulin (Cytoskeleton T240) to 5 mg/ml in 80 mM PIPES (pH 6.9), 2 mM MgCl$_2$, 0.5 mM EGTA, and 1 mM GTP; and rabbit skeletal myosin II (Cytoskeleton MY02) to 10 mg/ml in 25 mM PIPES (pH 7.0), 1.25 M KCl, 2.5% sucrose, 0.5% dextran, and 1 mM DTT. We flash freeze all proteins in experimental-sized aliquots and store at -80°C. We reconstitute the UV-sensitive myosin II inhibitor, (-)-blebbistatin (Sigma B0560) in anhydrous DMSO and store at -20°C for up to 6 months. Immediately prior to experiments, we remove enzymatically dead myosin II from aliquots using centrifugation clarification, as previously described[14,22].

**Composite Network Assembly:** We prepare actin-microtubule composites by mixing actin monomers, tubulin dimers and a trace amount of 1 µm diameter carboxylated microspheres (Polysciences) in PEM-100 (100 mM PIPES, 2 mM MgCl$_2$, and 2 mM EGTA), 0.1% Tween-20, 1 mM ATP, and 1 mM GTP, and incubating at 37°C for 30 minutes to allow for polymerization of actin filaments and microtubules. We coat microspheres (beads) with AlexaFluor594 BSA (Invitrogen) to visualize the particles and prevent nonspecific interactions with the composite[30,31]. We fix the combined molar concentration of actin and tubulin to $c = c_A + c_T = 5.8$ µM and the ratio of myosin to actin to $R = 0.08$, and vary the molar fraction of actin in the composite ($c_A/c = \phi_A$) from $\phi_A = 0$ to 1 in 0.25 increments (Fig 1A). To stabilize actin filaments and microtubules, we add an equimolar ratio of phalloidin to actin and a saturating concentration of Taxol (5 µM)[32,33]. We add an oxygen scavenging system (45 µg/ml glucose, 0.005% β-mercaptoethanol, 43 µg/ml glucose oxidase, and 7 µg/ml catalase) to inhibit photobleaching, and add 50 µM blebbistatin to control actomyosin activity.

**Sample Preparation and Imaging:** We pipet prepared composites into capillary tubing with an inner diameter of 800 µm, then seal with epoxy. Microspheres are imaged using a custom-built fLSM with a 10× 0.25NA Nikon Plan N excitation objective, a 20× 1.0 NA Olympus XLUMPlanFLN detection objective, and an Andor Zyla 4.2 CMOS camera[5]. A 561 nm laser is formed into a sheet to image the microspheres, while a collimated 405 nm laser is used to deactivate the blebbistatin, thereby activating actomyosin activity. Each acquisition location is at least 1 mm away from the previous one to ensure that there is no myosin activity when the image acquisition begins. For SPT, we collect ≥15 time-series consisting of ≥2000 frames, each with a 1000×300 pixel (194×58 µm$^2$) field of view (FOV), at 10 frames per second (fps). For DDM, we collect ≥3 time-series of ≥4000 frames, each with a 768×266 pixel (149×52 µm$^2$) FOV, at 10 fps.

**Single-Particle Tracking (SPT):** We use the Python package Trackpy[34] to track particle trajectories and measure the $x$- and $y$- displacements ($\Delta x, \Delta y$) of the beads as a function of lag times $\Delta t = 0.1$ s to 50 s. From the particle displacements, we use a custom-written Python script to calculate the time-averaged mean-squared displacement of the ensemble, $MSD = \frac{1}{2}(\Delta x^2 + \Delta y^2)$, from which we compute an anomalous scaling exponent, $\alpha$, via $MSD \sim \Delta t^\alpha$. Additionally, we compute van Hove probability distributions of particle displacements, $G(\Delta x, \Delta t)$ (Fig 1E), for 10 different lag times that span $\Delta t = 0.1$ s to 15 s. Following previous works[5–7], we fit each distribution for a given lag time to a sum of a Gaussian and exponential function: $G(\Delta x) = Ae^{-\frac{\Delta x^2}{2\sigma^2}} + Be^{-\frac{|\Delta x|}{\lambda}}$, where $A$ is the amplitude of the Gaussian term, $\sigma^2$ is the variance, $B$ is the amplitude of the exponential term, and $\lambda$ is the exponential decay constant.

**Differential Dynamic Microscopy (DDM):** We obtain the image structure function $D(q, \Delta t)$, where $q$ is the magnitude of the wave vector, following our previously described methods[12,35]. We fit each image structure function, or DDM matrix, to the following function:

$$D(q, \Delta t) = A(q)\left[1 - \exp\left[-(\Delta t/\tau(q))^{\gamma(q)}\right]\right] + B(q),$$

where $\tau(q)$ is the density fluctuation decay time, $\gamma$ is the stretching exponent, $A$ is the amplitude, and $B$ is the background[5,6]. $\tau(q)$ is a measure of the timescale over which particle density fluctuations decorrelate over a given lengthscale $\ell = 2\pi/q$. By fitting $\tau(q)$ to a power-law (i.e., $\tau(q) \sim q^{-\beta}$) we determine the dominant mode of transport, with $\beta = 2$, $>2$, and $<2$, indicating normal Brownian diffusion, subdiffusion and superdiffusion, respectively. We also examine the stretching exponent $\gamma$ that we extract from fitting $D(q, \Delta t)$ as another transport metric, with $\gamma < 1$ indicative of confined and heterogeneous dynamics[5,36–38] and $\gamma > 1$ indicative of active ballistic-like motion[14,39–41].

## 3 RESULTS AND DISCUSSION

To elucidate the combined effects of non-equilibrium activity and steric hindrance on particle transport in crowded active matter, we leverage the tunability of reconstituted cytoskeleton composites[42] and the power of coupling real-space (SPT) and reciprocal space (DDM) transport analysis, to robustly characterize particle transport as a function of active substrate content. Specifically, we tune the composition of actomyosin-microtubule composites to display a wide range of transport characteristics by varying the molar fraction of actomyosin, which we denote by the molar actin fraction $\phi_A$, keeping the myosin molarity fixed at 8% of $\phi_A$ (Fig 1A, Methods).

In Figure 2A, we plot the ensemble-averaged $MSD$ as a function of lag time $\Delta t$ for particles diffusing in composites of varying $\phi_A$. While $\phi_A = 0$ (no actomyosin) exhibits subdiffusive transport across the entire $\Delta t$ range, with $\alpha \simeq 0.67$, all $\phi_A > 0$ composites display biphasic transport which is subdiffusive ($\alpha < 1$) at short $\Delta t$ and superdiffusive ($\alpha > 1$) at long $\Delta t$. To more clearly show the transition from subdiffusion to superdiffusion, we plot $MSD/\Delta t$ versus $\Delta t$ (Fig 2B), which is a horizontal line for normal Brownian diffusion with the $\Delta t$-independent magnitude proportional to the diffusion coefficient. Positive and negative slopes correspond to superdiffusion and subdiffusion, respectively, with $MSD/\Delta t \sim \Delta t^1$ indicating ballistic motion. Corresponding $MSD/\Delta t$ magnitudes are proportional to $\Delta t$-dependent transport coefficients. While all $\phi_A > 0$ composites exhibit similar general trends with $\Delta t$, the lag time at which the dynamics transition from subdiffusive to superdiffusive, along with the degree to which $\alpha$ deviates from $\alpha = 1$ in each regime, depend non-trivially on $\phi_A$. Moreover, as clearly seen in Fig 2C, while $\alpha$ values for active composites ($\phi_A > 0$) transition to superdiffusive at longer lag times, the magnitudes of the transport coefficients remain nearly an order of magnitude smaller than that of the inactive network ($\phi_A = 0$) at any given $\Delta t$.

To evaluate the $\phi_A$-dependence of the biphasic behavior, we first compute the lag times at which each composite transitions out of subdiffusive scaling, denoted as $\Delta t_1$, and transitions into superdiffusive scaling, denoted as $\Delta t_2$. Note that for some composites there is an extended plateau regime between the two timescales such that $\Delta t_1$ and $\Delta t_2$ are substantially separated. To quantify $\Delta t_1$, we compute the largest lag time for which linear regression of $\log MSD$ versus $\log \Delta t$ over the range $\Delta t \in [0.1 \text{ s}, \Delta t_1]$ yields $R^2 > 0.99$. Likewise, we define $\Delta t_2$ as the shortest $\Delta t$ for which the same linear regression yields $R^2 > 0.99$ for the range $\Delta t \in [\Delta t_2, 50 \text{ s}]$ (Fig 2C). We find that both timescales generally decrease with increasing $\phi_A$ as does the separation between the timescales ($\Delta t_2 - \Delta t_1$), suggesting that the rate of particle motion mediated by directed motor-driven network dynamics increases with increasing $\phi_A$. In other words, the time it takes for the active dynamics to be 'felt' by the particles, driving them out of their confined subdiffusive motion, decreases with increasing $\phi_A$.

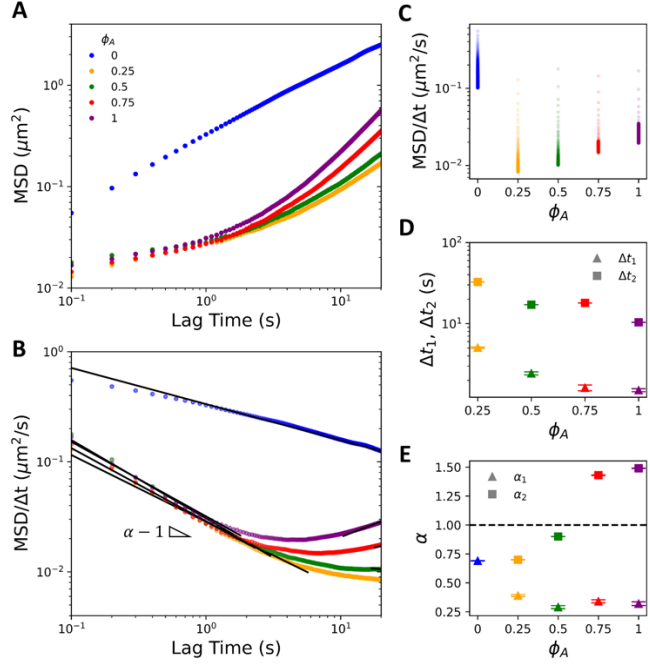

**Figure 2: Actomyosin activity in actin-microtubule composites endows biphasic particle transport marked by pronounced subdiffusion at short lag times and superdiffusion at long lag times.** (**A**) Mean-squared displacements ($MSD$) plotted versus lag time $\Delta t$ for myosin-driven actin-microtubule composites with molar actin fractions of $\phi_A = 0$ (blue), 0.25 (gold), 0.50 (green), 0.75 (red), and 1 (purple). Fits of the data to $MSD \sim \Delta t^\alpha$, shown in (B), determine the anomalous scaling exponent $\alpha$ that describes the dynamics (see Fig 1). (**B**) Mean-squared displacements scaled by lag time ($MSD/\Delta t$) plotted versus lag time $\Delta t$ delineate regions of subdiffusion (negative slopes) and superdiffusion (positive slopes). Color coding is according to the legend in A. Black lines indicate fits to $MSD \sim \Delta t^\alpha$ over the short ($\Delta t < \Delta t_1$) and long ($\Delta t > \Delta t_2$) time regimes where each curve is well-fit by a single power law. (**C**) Data shown in B plotted versus actin fraction $\phi_A$, with all $MSD/\Delta t$ values for each $\phi_A$ plotted along the same vertical, with the gradient indicating increasing $\Delta t$ from light to dark. The magnitude of each data point is proportional to a transport rate, with higher values indicating faster motion. (**D**) Lag times at which each composite transitions out of subdiffusive transport ($\Delta t_1$) and transitions into superdiffusivity ($\Delta t_2$). (**E**) Anomalous scaling exponent $\alpha$ derived from fits shown in (B) for $\Delta t < \Delta t_1$ ($\alpha_1$) and $\Delta t > \Delta t_2$ ($\alpha_2$). Dashed line at $\alpha = 1$ represents scaling indicative of normal Brownian diffusion. Values above and below the line indicate superdiffusion and subdiffusion, respectively. For both (C) and (D) error bars indicate standard error of the mean. Color-coding in all subfigures matches the legend in A.

To understand this phenomenon, we consider that active ballistic transport would only be detectable at timescales in which the network motion can move a bead more than the minimum resolvable displacement: $\Delta t_a \approx (100 \text{ nm})/(network\ speed)$. Using reported speed values of $v \approx 2.2 - 85$ nm/s for similar myosin-driven composites [23], we compute $\Delta t_a \approx (100 \text{ nm})/v \approx 1 - 50$ s, aligning with our $\Delta t_1$ and $\Delta t_2$ values, and thus corroborating that the deviation from sub-diffusion and transition to superdiffusion is due to myosin-driven ballistic motion. Moreover, the previously reported speeds generally decreased with decreasing $\phi_A$, such that $\Delta t_a$ should increase to values that are beyond our experimental range for $\phi_A < 0.75$, just as we see in Fig 2B in which $\phi_A < 0.75$ composites do not display a clear uptick to superdiffusive dynamics.

To determine the extent to which motor-driven transport and confinement contribute to the particle dynamics, we next evaluate the anomalous scaling exponent in the short and long $\Delta t$ regimes by performing power-law fits to the $MSD$s in each regime (Figure 2D). Surprisingly, the scaling exponents in the $\Delta t < \Delta t_1$ regime for all active composites ($\phi_A > 0$) are markedly smaller (more subdiffusive) than the inactive composite ($\phi_A = 0$), with $\phi_A$-dependent values of $\alpha \simeq 0.29 - 0.39$ compared to $\alpha \simeq 0.67$ for the $\phi_A = 0$ network. To understand the decrease in $\alpha$ with increasing $\phi_A$ for the active composites, as well as the unexpected ~2-fold reduction in $\alpha$ for active composites, we turn to previous studies[5,6], that reported that, in the absence of any crosslinking, steady-state actin-microtubule composites exhibit subdiffusion with scaling exponents that decrease from $\alpha \approx 0.82$ to $\alpha \approx 0.56$ as $\phi_A$ increases from 0 to 1. This monotonic ~30% decrease with increasing $\phi_A$, similar to the ~25% decrease we observe with increasing $\phi_A$, was suggested to arise from increased composite mobility that entrains the bead motion as rigid microtubules are replaced with semiflexible actin filaments[6,12]. This mobility is paired with a decreasing mesh size as $\phi_A$ increases, which, in turn, increases composite viscoelasticity and particle confinement, both of which contribute to decreasing $\alpha$[6].

To understand the lower $\alpha$ values we measure, compared to those previously reported for steady-state composites, we look to previous studies on $\phi_A = 0.5$ actin-microtubule composites with varying types of static crosslinking. In these studies, subdiffusion is much more extreme ($\alpha \approx 0.33$) when actin filaments are crosslinked to each other versus when there is no crosslinking ($\alpha \approx 0.64$) [5]. Taken together, our results suggest that the ~2-fold reduction in $\alpha$ between $\phi_A = 0$ and $\phi_A > 0$ composites likely arises from myosin motors acting as static crosslinkers on timescales shorter than the timescale over which they can actively translate the composite. As described above, myosin acting as a static crosslinker for $\Delta t < \Delta t_1$ is consistent with previously reported speeds for myosin-driven composites[5,6], as well as reported actomyosin turnover rates[24]. The weak decrease in $\alpha$ with increasing $\phi_A$ likely arises from the decreasing mesh size and increasing mobility of the network as $\phi_A$ increases[43], as described above.

Finally, examining the long-time regime, $\Delta t > \Delta t_2$, our results show that higher actomyosin fractions correspond to higher $\alpha$ values, increasing ~2-fold from ~0.73 for $\phi_A = 0.25$ to ~1.47 for $\phi_A = 1$. Moreover, only $\phi_A > 0.5$ composites exhibit an uptick to superdiffusive dynamics ($\alpha > 1$) over our measurement window, suggesting that the extent to which myosin-driven dynamics contribute to particle transport scales with the fraction of active substrate. Moreover, the timescale at which its contribution dominates particle transport is determined by the network speed, which increases with increasing $\phi_A$ as described above[23].

To shed further light on the mechanisms underlying the anomalous transport shown in Fig 2, we compute van Hove distributions $G(\Delta x, \Delta t)$ for two decades of lag times ($\Delta t = 0.1 - 15\ s$) (Fig 3A). From the distributions, we first compute anomalous scaling exponents $\alpha$, to corroborate our $MSD$ analysis, by recalling that the full width at half maximum, $FWHM$, for a Gaussian distribution scales with the standard deviation $\sigma$ as $FWHM = 2\sqrt{2ln2}\ \sigma$. Because $\sigma^2 \sim \Delta x^2$, by definition, and $MSD \sim \Delta x^2 \sim \Delta t^\alpha$, we determine $\alpha$ by computing $FWHM$ for each distribution and fitting the $\Delta t$-dependent values to the power-law $FWHM(\Delta t) \sim (\Delta t)^{\alpha/2}$ (Fig 3B)[7,44]. As shown in Figs 3B,C, $FWHM(\Delta t)$ for $\phi_A = 0$ fits well to a single power-law, with $\alpha \simeq 0.7$, nearly indistinguishable from that computed from the $MSD$, across the entire $\Delta t$ range. Conversely, informed by the biphasic $MSD$ scaling we observe for active composites (Fig 2C), we fit $FWHM(\Delta t)$ for each active composite to separate power-law functions over short ($0.15\ s < \Delta t < 2\ s$) and long ($2\ s < \Delta t < 15\ s$) lag times, relative to the average $\Delta t_1$ we determine from $MSD$ fits. Further, similar to the $\phi_A$-dependence of $\alpha_1$ and $\alpha_2$ values determined from $MSD$s, the scaling exponents determined from $FWHM$, increase with increasing $\phi_A$, with $\alpha_1$ (for $\Delta t < 2\ s$) increasing from ~0.62 to ~0.85, similar to values reported for steady-state actin-microtubule composites[6], and $\alpha_2$ (for $\Delta t > 2\ s$) spanning from subdiffusive to superdiffusive. Higher $\alpha_1$ values compared to those determined via $MSD$s are likely due to the lower $\Delta t$ resolution in $FWHM$ fitting and the single $\Delta t$ value used to divide the two regimes.

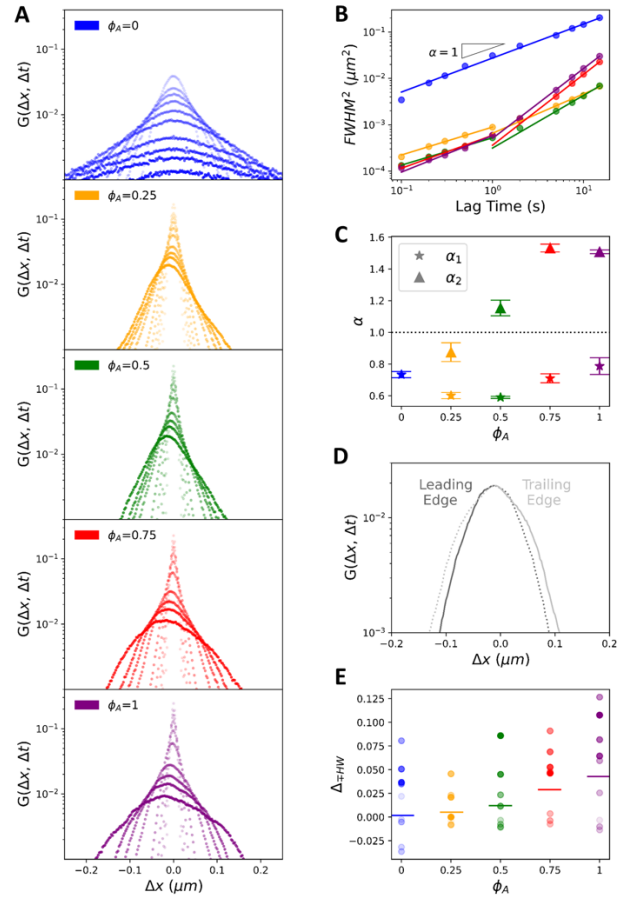

**Figure 3: Asymmetric non-Gaussian van Hove distributions reveal a combination of heterogeneous subdiffusion and advective transport of particles in active composites.** (**A**) van Hove distributions $G(\Delta x, \Delta t)$ of particle displacements $\Delta x$, measured via SPT, for lag times $\Delta t = 0.1, 0.2, 0.3, 0.5, 1, 2, 3, 5, 10, 15\ s$ denoted by the color gradient going from light to dark for increasing $\Delta t$. Each panel corresponds to a different composite demarked by their $\phi_A$ value with color-coding as in Fig 2. (**B**) The square of the full width at half-maximum $(FWHM)^2$ versus lag time $\Delta t$ for each composite shown in A. Solid lines are fits to $(FWHM)^2 \sim \Delta t^\alpha$. For $\phi_A > 0$ composites we fit short ($\Delta t \leq 1$ s) and long ($\Delta t \geq 1$ s) lag time regimes separately. (**C**) The scaling exponents $\alpha$ as functions of $\phi_A$ determined from the fits shown in B, where $\alpha_1$ (stars) and $\alpha_2$ (triangles) correspond to scalings for the short and long $\Delta t$ regimes, respectively. The dashed horizontal line denotes scaling for normal Brownian diffusion. (**D**) A sample $G(\Delta x, \Delta t)$ distribution ($\phi_A = 0.75$ at $\Delta t = 10s$) showing the asymmetry about the mode value $\Delta x_{peak}$. We divide each distribution into a leading edge (dark grey, displacements of the same sign as $\Delta x_{peak}$ and greater in magnitude) and the trailing edge (light grey, the remaining part of the distribution). To clearly demonstrate the asymmetry, we mirror each edge about $\Delta x_{peak}$ using dashed lines. (**E**) The fractional difference of the half-width at half maximum $HWHM$ of the trailing (-) edge from the leading (+) edge, ($\Delta_{\mp HW} = HWHM_- - HWHM_+) / HWHM_+$), for each $\phi_A$ and $\Delta t$. Color coding and gradient indicate $\phi_A$ and $\Delta t$, respectively, as in A. Horizontal bars through each distribution denote the mean.

While our analysis described above assumes Gaussian distributions, Fig 3A shows that nearly all distributions have distinct non-Gaussian features similar to those reported for steady-state actin-microtubule composites[5,7]. In particular, $G(\Delta x, \Delta t)$ distributions for the inactive network ($\phi_A = 0$) exhibit pronounced exponential tails at large displacements. This non-Gaussianity, seen in other crowded and confined soft matter systems[7], is a signature of heterogeneous transport and can also indicate caging and hopping between cages.

The distributions for active composites are even more complex, with asymmetries and peaks at $\Delta x \neq 0$ (Fig 3A), not readily predictable from our $MSD$ analysis. The first interesting feature we investigate is the non-zero mode value $\Delta x_{peak}$ that increases in magnitude with increasing $\Delta t$, indicating directed ballistic-like motion, thereby corroborating our superdiffusive scaling exponents. Perhaps less intuitive is the robust asymmetry between the 'leading (+) edge' and 'trailing (-) edge' of each distribution, which we define by splitting each distribution about its peak, $\Delta x_{peak}$. Specifically, the leading edge is the part of the distribution that has displacements of the same sign as $\Delta x_{peak}$ and greater in magnitude, while the remaining part is the trailing edge (Figure 3D). We observe that for most distributions the leading edge appears more Gaussian-like while the trailing edge exhibits pronounced large-displacement 'tails'. To broadly quantify this asymmetry, we evaluate the half-width at half-maximum ($HWHM$) for the leading (+) and trailing (-) edges of each distribution and compute the percentage increase in $HWHM$ for the trailing versus leading edge: $\Delta_{\mp HW} = (HWHM_- - HWHM_+)/HWHM_+$ (Fig 3E). We find that $\Delta_{\mp HW}$ is positive for all active composites and increases with increasing $\phi_A$, demonstrating that the asymmetry is a direct result of active composite dynamics which contribute more to the transport as the actomyosin content increases.

To more quantitatively characterize the rich transport phenomena revealed in Fig 3, we first fit each $G(\Delta x, \Delta t)$ to a sum of a Gaussian and an exponential (see Methods), as done for steady-state cytoskeleton composites[5–7]. Fig 4A compares the distributions and their fits for all composite formulations at $\Delta t = 0.3\ s$ (top panel) and $10\ s$ (bottom panel), and Fig 4B displays zoom-ins of the corresponding leading and trailing edges. As shown, while this sum describes the inactive network distributions reasonably well, it overestimates leading edge displacements and underestimates trailing edge displacements of the active networks (Fig 4B). This asymmetry suggests that the leading edges are more Gaussian-like and the trailing edges are more exponential-like. To account for this asymmetry, we fit each half of each distribution separately to a one-sided sum of a Gaussian and exponential and evaluate the relative contributions from the Gaussian and exponential terms. As detailed in the Methods, we denote the amplitude of the Gaussian term and exponential term as $A$ and $B$, respectively, such that their relative contributions are $a = A/(A + B)$ and $b = B/(A + B)$.

As shown in Fig 4C,D, in which $a$ and $b$ are normalized by the corresponding $\phi_A = 0$ value and plotted for each $\phi_A$, active composites are more Gaussian-like ($a/a(\phi_A = 0) > 1$) and less exponential ($b/b(\phi_A = 0) > 1$) than the inactive system for both leading and trailing edges, suggesting that the active processes that induce contraction and flow of the composites, likewise reduce transport heterogeneity and intermittent hopping, possibly by promoting mixing and advection. Consistent with this interpretation is the observation that the Gaussian contribution $a$ increases with increasing $\phi_A$ and is consistently larger for the leading edge, which consists of displacements oriented with the direction of the myosin-driven composite motion (Fig 4D).

Conversely, the increased contribution from the exponential term for the trailing edge, implies that displacements comprising these exponential tails are dominated by heterogeneous transport modes such as hopping between heterogeneously distributed cages [6,7]. To better understand this effect, we recall that particle displacements comprising the trailing edge are lagging behind the bulk translational motion of the composite. As the composite moves and restructures, a fraction of the particles remain caged in the moving composite and thus move along with it, corresponding to displacements comprising the leading edge, whereas a fraction of the particles are squeezed out or hop out of composite cages and into new 'trailing' cages. It is also likely that as the composite contracts and forms more heterogeneous structures and larger open voids that characteristic 'hopping' displacements, as well as displacements withing cages, may increase and become more heterogeneous, thereby enhancing exponential tails.

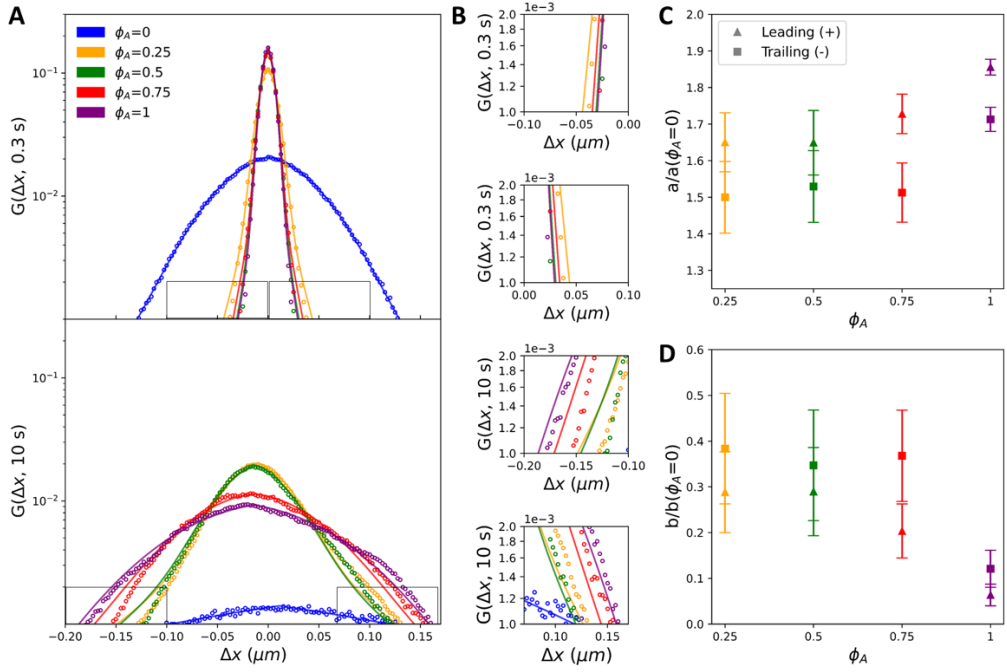

**Figure 4: Actomyosin activity reduces heterogeneous non-Gaussian diffusivity and endows Gaussian-like advective transport.** (**A**) Comparing van Hove distributions of composites with different $\phi_A$ (see legend) at lag times of $\Delta t = 0.3\ s$ (top) and $\Delta t = 10\ s$ (bottom). Color-coded solid lines are fits of each distribution to the sum of a Gaussian and an exponential: $G(\Delta x) = Ae^{-\frac{\Delta x^2}{2\sigma^2}} + Be^{-\frac{|\Delta x|}{\lambda}}$. Black rectangles indicate regions of the distributions that are shown zoomed-in in (**B**). (**C**) Fractional amplitude of the Gaussian term in each fit, $a = A/(A+B)$, normalized by the corresponding value for $\phi_A = 0$. Fits are performed separately for the leading (+, triangles) and trailing (-, squares) edges of each distribution. Data shown are the averages and standard deviations across all lag times for each $\phi_A$. (**D**) Fractional amplitude of the exponential term in each fit, $b = A/(A+B)$, normalized by the corresponding value for $\phi_A = 0$. Fits are performed separately for the leading (+, triangles) and trailing (-, squares) edges of each distribution. Data shown are the averages and standard deviations across all lag times for each $\phi_A$.

To expand the range of length and time scales over which we probe the non-equilibrium transport, and provide an independent measure of the dynamics, we complement our real-space SPT analysis with Fourier-space DDM analysis, as described in the Methods and previously [5,6,14]. Briefly, we compute the radially-averaged image structure function $D(q, \Delta t)$ of the Fourier transform of image differences as a function of wavevector $q$ and lag time $\Delta t$. From fits of $D(q, \Delta t)$ to a function with a stretched exponential term (see Methods, Fig 5A), we determine the $q$-dependent characteristic decay time $\tau(q)$ and stretching exponent $\gamma$ for each composite (Fig 5), which characterize the dynamics. $\tau(q)$ typically exhibits power-law scaling $\tau(q) \sim q^{-\beta}$ where $\beta$ is related to the anomalous scaling exponent $\alpha$ via $\beta = 2/\alpha$, such that $\beta > 2, \beta = 2, \beta < 2$ and $\beta = 1$ correspond to, respectively, subdiffusive, normal diffusive, superdiffusive, and ballistic motion. Similarly, stretching exponents $\gamma$ are typically 1 for normal Brownian motion, while $\gamma < 1$ is a signature of crowded and confined systems[15,40] and $\gamma > 1$ indicates active transport[15,45].

As shown in Fig 5B, $\tau(q)$ curves for all active composites follow scaling indicative of superdiffusive or ballistic transport while the $\phi_A = 0$ system more closely follows diffusive scaling. Further, $\tau(q)$ for $\phi_A = 0$ is an order of magnitude lower than for all active composites, indicating that particle transport is faster for the inactive composite, in line with our results shown in Fig 2C, despite the displacements exhibiting diffusive rather than ballistic-like motion. This effect can be more clearly seen in Fig 5D, which displays the $q$-dependent distribution of $\tau(q)^{-1}$ values, a measure of dynamic decorrelation rates, for each $\phi_A$. As shown, $\tau^{-1}$ values for $\phi_A = 0$ are an order of magnitude larger than those for

$\phi_A > 0$ composites. Fig 5D also shows that decorrelation rates in active composites increase modestly with increasing $\phi_A$ suggesting that transport is dictated primarily by active restructuring and flow, rather than crowding and confinement, which increases as actomyosin content increases. The lack of subdiffusive scaling or crossovers from sub- to super-diffusive dynamics for active composites (as our SPT analysis shows) can be understood as arising from the larger length and time scales DDM probes. Namely, DDM spans lengthscales of $2\pi/q \simeq 1.6 - 6.28$ μm and timescales of $\tau \simeq 20 - 100\ s$ (Fig 5A) compared to the ~$0.1 - 1.5$ μm and ~$0.1 - 50\ s$ length and timescales accessible to SPT.

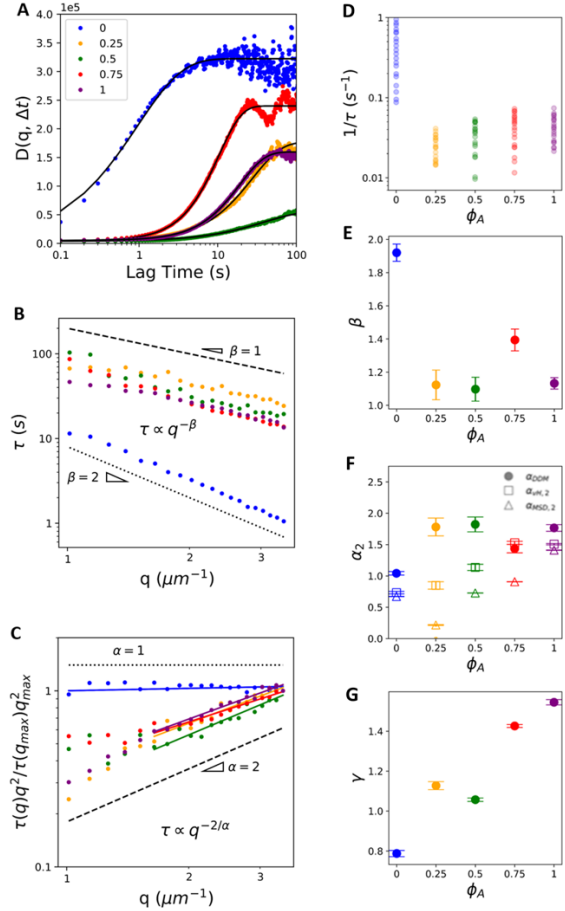

**Figure 5: DDM analysis reveals ballistic-like transport of particles entrained in active composites at mesoscopic spatiotemporal scales.** (**A**) Sample image structure functions $D(q, \Delta t)$ for composites with actin fractions $\phi_A$ indicated in the legend. All curves shown are evaluated at $q = 3.92\ \mu m^{-1}$, and solid black lines are fits to the data to determine corresponding $q$-dependent decay times $\tau(q)$ and stretching exponents $\gamma$, as described in Methods. (**B**) Decay times $\tau(q)$ for each composite shown in (A). Dashed and dotted black lines show scaling $\tau(q) \sim q^{-\beta}$ for ballistic ($\beta = 1$) and diffusive ($\beta = 2$) transport. (**C**) $\tau(q) \times q^2$, normalized by $\tau(q_{max}) \times (q_{max})^2$, for the data shown in (B). Horizontal dotted line and unity-sloped dashed line correspond to scaling indicative of normal diffusion ($\alpha = 2/\beta = 1$) and ballistic motion ($\alpha = 2/\beta = 2$). Color-coded solid lines correspond to power-law fits, with the corresponding exponents $\beta$ and $\alpha$ shown in (E) and (F). For $\phi_A > 0$ composites, the fitting range is truncated to $q > 1.5\ \mu m^{-1}$ where a single power-law is observed. (**D**) Scatter plot of $1/\tau(q)$, a measure of the transport rate, for all measured $q$ values for each $\phi_A$. Color coding and gradient indicate $\phi_A$ and $q$, respectively, with light to dark shades of each color indicating increasing $q$ values. (**E**) DDM scaling exponents $\beta$ determined from fits shown in (C). (**F**) Anomalous scaling exponents $\alpha_2$ determined from $\tau(q)$ fits (filled circles, $\alpha_{DDM} = 2/\beta$), as well as the large-$\Delta t$ regime fits of the $MSD$s (open triangles, $\alpha_{MSD,2}$) and van Hove distributions (open triangles, $\alpha_{vH,2}$) measured via SPT (see Figs 2,3). Error bars indicate 95% confidence intervals of fits. (**G**) Stretching exponent $\gamma(q)$, averaged over all $q$ values, for each composite $\phi_A$, with error bars indicating standard error.

To better visualize differences in $\tau(q)$ scaling between composites we plot $\tau(q) \times q^2$ normalized by $\tau(q_{max}) \times (q_{max})^2$ (Fig 5C). Diffusive transport manifests as a horizontal line, as we see for $\phi_A = 0$, while ballistic-like motion follows a power-law scaling of 1, which roughly describes the $\phi_A > 0$ curves. To quantify the DDM scaling exponent $\beta$ that describes the dynamics, we fit each $\tau(q)$ curve to a power-law (i.e., $\tau(q) \sim q^{-\beta}$) (Fig 5C,E). For the active composites, we restrict our fitting range to $q > 1.5$ μm$^{-1}$, in which a single power-law is observed. For smaller $q$ values (larger length and time scales), we note that $\phi_A = 0.25$ and 1 composites exhibit roughly ballistic motion whereas $\phi_A = 0.5$ and 0.75 exhibit roughly diffusive dynamics (Fig 5B)[15,16]. However, we restrict further quantification and interpretation of this small-$q$ regime as it comprises relatively few data points and low statistics. Over the range that we fit our data, we find that $\beta \simeq 1.92$ for the inactive composite, indicative of diffusive dynamics, whereas active composites exhibit near-ballistic values of $\beta \simeq 1.03 - 1.26$. To directly compare $\beta$ values to the anomalous scaling exponents $\alpha$ that we determine from SPT (Figs 2E, 3B), we plot $\alpha_{DDM} = 2/\beta$ (Fig 5C,F) with the $\alpha$ values we determined from the $MSD$s and van Hove distributions in the large $\Delta t$ regime, which we denote as $\alpha_{MSD,2}$ and $\alpha_{vH,2}$. Scaling exponents determined from all three methods follow similar trends with $\phi_A$ with active composites displaying larger $\alpha$ values than the $\phi_A = 0$ system. Generally, for each $\phi_A$, we find that $\alpha_{DDM} > \alpha_{vH,2} > \alpha_{MSD,2}$, which arises from the different timescales probed by each method. Namely, all systems tend to subdiffusion at short lag times (measured most accurately via $MSD$s) and free diffusion or ballistic motion at large lag times (accessed only by DDM), so scaling exponents measured at short lag times should generally be lower than those measured over larger lag times.

Finally, to shed light on the competing contributions from motor-driven dynamics versus confinement and crowding to transport at larger spatiotemporal scales, we evaluate the dependence of the stretching exponent $\gamma$ on $\phi_A$. Fig 5G shows that transport in the inactive network is described by $\gamma \simeq 0.79$, indicating that confinement dominates over active dynamics (i.e., $\gamma < 1$), whereas all $\phi_A > 0$ composites exhibit $\gamma > 1$, indicative of transport governed largely by active dynamics. Moreover, $\gamma$ generally increases as the actomyosin fraction increases, corroborating the dominant role that active composite dynamics plays in the rich transport phenomena we reveal[15].

## 4 CONCLUSIONS

Here, we couple real-space SPT and Fourier-space DDM to characterize particle transport across three decades in time (~10$^{-1}$ - 10$^2$ s) and two decades in space (~10$^{-1}$ - 10 μm) in biomimetic composites that exhibit both pronounced crowding and confinement as well as active motor-driven restructuring and flow. Using our robust approach, we discover and dissect novel transport properties that arise from the complex interplay between increasing activity and confinement as the actomyosin fraction increases. Myosin motors induce ballistic-like contraction, restructuring and flow of the composites, leading entrained particles to exhibit similar superdiffusive, advective and Gaussian-like transport. Conversely, steric entanglements, connectivity and slow thermal relaxation of cytoskeletal filaments mediate heterogeneous, subdiffusive transport of confined particles.

Figure 6 summarizes and compares the key metrics we present in Figs 2-5 that characterize these complex transport properties. Importantly, as highlighted in Figure 6, while there is clear difference between the inactive and active networks for nearly all of the transport metrics we present, we emphasize that there are very few clear monotonic dependences on $\phi_A$ for the active composites. This non-monotonic complexity is a direct result of the competition between motor-driven active dynamics, crowding and connectivity – all of which increase with increasing actomyosin content. These intriguing transport characteristics have direct implications in key cellular processes in which actomyosin and microtubules synergistically interact, such as cell migration, wound healing, cytokinesis, polarization and mechano-sensing[23]. Moreover, our robust measurement and analysis toolbox and tunable active matter platform, along with the complex transport phenomena we present, are broadly applicable to a wide range of active matter and biomimetic systems of current intense investigation.

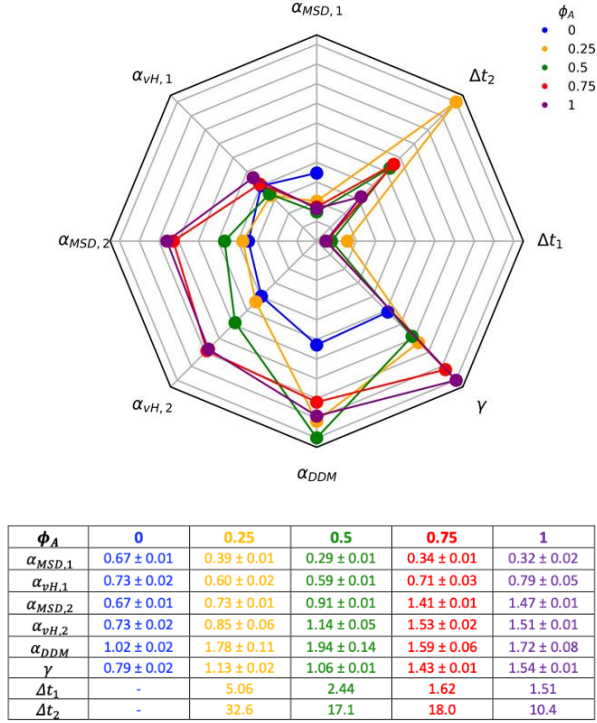

| $\phi_A$ | 0 | 0.25 | 0.5 | 0.75 | 1 |
|---|---|---|---|---|---|
| $\alpha_{MSD,1}$ | 0.67 ± 0.01 | 0.39 ± 0.01 | 0.29 ± 0.01 | 0.34 ± 0.01 | 0.32 ± 0.02 |
| $\alpha_{vH,1}$ | 0.73 ± 0.02 | 0.60 ± 0.02 | 0.59 ± 0.01 | 0.71 ± 0.03 | 0.79 ± 0.05 |
| $\alpha_{MSD,2}$ | 0.67 ± 0.01 | 0.73 ± 0.01 | 0.91 ± 0.01 | 1.41 ± 0.01 | 1.47 ± 0.01 |
| $\alpha_{vH,2}$ | 0.73 ± 0.02 | 0.85 ± 0.06 | 1.14 ± 0.05 | 1.53 ± 0.02 | 1.51 ± 0.01 |
| $\alpha_{DDM}$ | 1.02 ± 0.02 | 1.78 ± 0.11 | 1.94 ± 0.14 | 1.59 ± 0.06 | 1.72 ± 0.08 |
| $\gamma$ | 0.79 ± 0.02 | 1.13 ± 0.02 | 1.06 ± 0.01 | 1.43 ± 0.01 | 1.54 ± 0.01 |
| $\Delta t_1$ | - | 5.06 | 2.44 | 1.62 | 1.51 |
| $\Delta t_2$ | - | 32.6 | 17.1 | 18.0 | 10.4 |

**Figure 6: A robust suite of metrics reveals complex scale-dependent transport resulting from competition between motor-driven active dynamics, crowding and network connectivity.** The 8-variable spider plot shows how the key metrics we use to characterize transport depend on $\phi_A$ (color-code shown in legend). A greater distance from the center signifies a larger magnitude. $\alpha$ values determined from DDM ($\alpha_{DDM}$), SPT $MSDs$ ($\alpha_{MSD,1}$, $\alpha_{MSD,2}$, $\alpha_{vH,1}$, $\alpha_{vH,2}$ and SPT van Hove distributions ($\alpha_{vH,1}$, $\alpha_{vH,2}$) are scaled identically for direct comparison, as are the two timescales determined from $MSDs$ ($\Delta t_1$, $\Delta t_2$). The stretching exponent $\gamma$ is scaled independently. The table provides the values with error for each metric plotted.

**Competing Interests**

The authors declare no competing interests.

## DATA AVAILABILITY STATEMENT

Data will be available by contacting the corresponding author Janet Y Sheung.

## AUTHOR CONTRIBUTIONS

JYS analyzed and interpreted the data and wrote the paper; JG collected and helped analyze the data; SKK and BYL helped analyze data and prepare figures; RJM conceived and guided the project, developed analysis code, helped analyze and interpret the data, and helped write the paper; RMRA conceived and guided the project, interpreted the data, and wrote the paper.

**FUNDING**

This work was supported by National Institutes of Health R15 Awards (R15GM123420, 2R15GM123420-02) to RMR-A and RJM, and William M. Keck Foundation Research Grant to RMR-A. JYS acknowledges startup funding from Scripps, Pitzer, and Claremont McKenna Colleges.

**ACKNOWLEDGMENTS**

We thank Aaron Xie, Nadia Schwartz Bolef, and Jemma Kushen for help with data analysis and visualization. We thank Gloria Lee for help with composite design. We thank Daisy Achiriloaie for insightful discussions.